\documentclass{aa}
\usepackage{psfig} 
\begin{document}

\title{BeppoSAX observations of the Narrow-Line Seyfert 1 galaxy\\  
\object{RX J1702.5+3247}}
  
\author{M. Gliozzi\inst{1}  
\and  W. Brinkmann\inst{1} 
\and S. A. Laurent-Muehleisen\inst{2}\thanks{Visiting Astronomer, Kitt Peak 
National Observatory, National Optical
Astronomy Observatories, which is operated by the Association of Universities
for Research in Astronomy, Inc. (AURA) under cooperative agreement with the
National Science Foundation}
\and E.C. Moran\inst{3}
\and J. Whalen\inst{2} } 
\offprints{mgliozzi@xray.mpe.mpg.de} 
\institute{
Max-Planck-Institut f\"ur extraterrestrische Physik,
         Postfach 1312, D-85741 Garching, Germany
\and  Department of Physics, University of California - Davis, 1 Shields Ave.,
Davis, CA, 95616 , USA
\and Department of Astronomy, University of California, Berkeley, CA 94720 USA}

\date{Received: ; accepted: }

\abstract{
We report optical, radio and X-ray observations of the 
Narrow-Line Seyfert 1 galaxy  \object{RX J1702.5+3247}.
The soft (0.1-2 keV) X-ray flux, measured by BeppoSAX, is characterized by
strong variability on short time scales ($< 500$  sec). 
The most extreme 
amplitude variations require a radiative efficiency exceeding the maximum
for a Kerr black hole, implying the presence of relativistic effects.
A comparison with archival ROSAT data reveals long term
temporal and spectral variability. The 0.1-10 keV spectrum is equally well
fitted either by an ionized reflection disk model, 
or by a broken power law plus a Gaussian line, consistent with a
hydrogen--like iron line at 6.97 keV from a highly ionized accretion disk. 
\keywords{Galaxies: active -- 
Galaxies: fundamental parameters  
-- Galaxies: nuclei -- X-rays: galaxies }
}
\titlerunning{BeppoSAX observations of \object{RX J1702.5+3247}}
\authorrunning{M.~Gliozzi et al.}
\maketitle
\section{Introduction}
Narrow-line Seyfert 1 galaxies (NLS1; see Osterbrock \& Pogge \cite{oster-pog})
are a peculiar class of AGN defined by their optical line properties:
the H$\beta$ FWHM does not
exceed 2000 km~s$^{-1}$, the [O~III]$\lambda$5007 to H$\beta$ ratio is
less than 3, and the UV$-$visual spectrum is usually rich in
high-ionization lines and strong Fe~II emission multiplets.

ROSAT and ASCA observations have shown that the  X-ray properties of
NLS1 galaxies are remarkably different from those of `normal' Seyfert 1 
galaxies.
Large amplitude
X-ray variability with time scales as short as few hundred seconds 
(Forster \& Halpern \cite{forst}, Boller \cite{boll2}, Brandt \cite{brand2})
and very steep soft X-ray spectra (Boller \cite{boll1})  
are common among these objects. On average, NLS1 galaxies have steeper 
energy spectra 
also in the hard 2-10 keV energy band (Brandt \cite{brand1}) and show higher
variability (Fiore \cite{fior}, Leighly \cite{leig1}).
A clear anti-correlation between the soft
X-ray spectral slope and the H$\beta$ FWHM has been found over a wide range
of X-ray luminosities (Boller \cite{boll1}, Laor \cite{laor}). 
Emission lines of highly ionized iron have been discovered in several NLS1
galaxies  (Pounds \cite{poun1},
Comastri \cite{coma1}, Turner \cite{turn}, Leighly \cite{leig2}). 
This supports the hypothesis of
the presence of a particularly high accretion rate in NLS1 galaxies
(Pounds \cite{poun1}), as the ionization parameter scales with
$(\dot M/\dot M_{\rm Edd})^3$ (Ross \& Fabian \cite{ros-fab}).

NLS1 are rarely radio-loud (Remillard \cite{remi}, Ulvestad \cite{ulve},
Siebert \cite{sibe}, Grupe \cite{grupe}), and thus   
their radio properties have been poorly explored.
An unexpectedly large number of NLS1 has been discovered among the sources
found in the cross correlation of the ROSAT All Sky Survey
(RASS) and the FIRST VLA radio survey (Brinkmann \cite{brink}, Becker \cite
{beck}).
\begin{figure*}
\psfig{figure=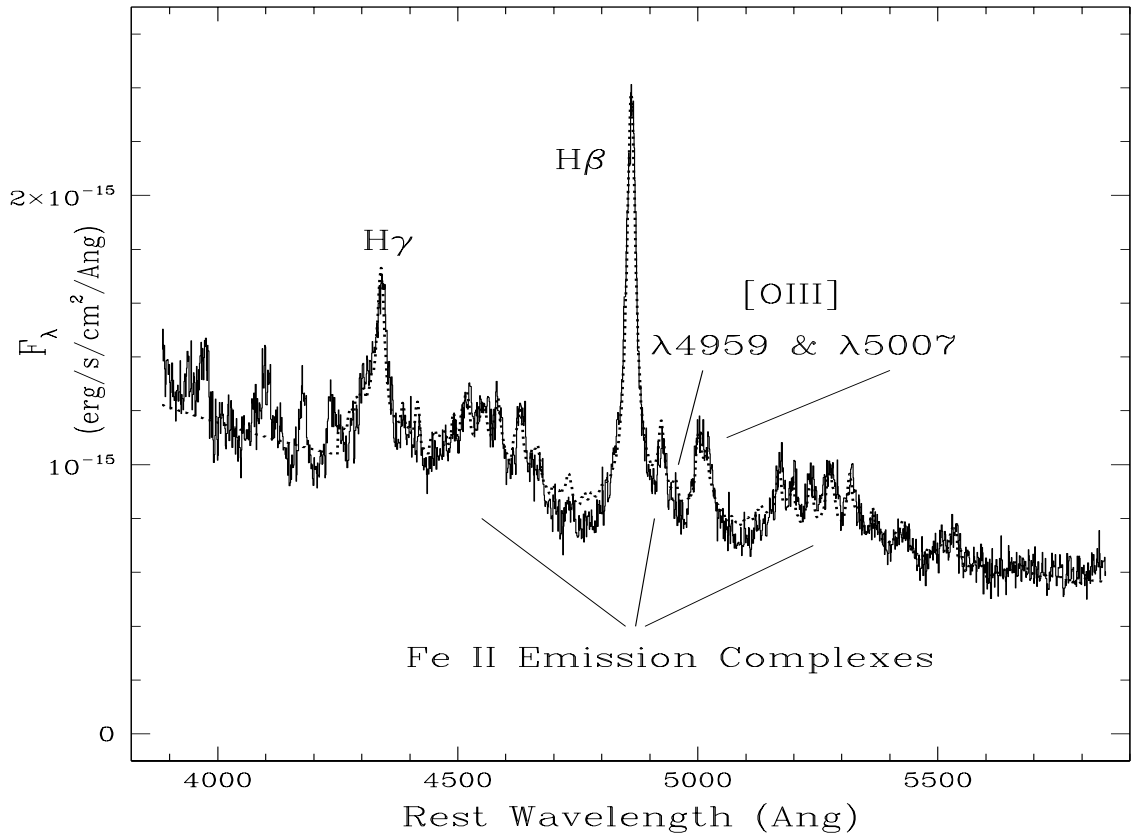,height=11cm,width=15.cm,%
bbllx=19pt,bblly=160pt,bburx=565pt,bbury=690pt,angle=0,clip=}
\caption{A blow up of the 3880 - 5850 \AA ~region in the observed broad band 
spectrum of \object{RX J1702.5+3247}.
The spectrum was taken on 4 June, 1997, and has a resolution of
3.7 \AA.  Because observing conditions were not photometric, the flux scale 
shown on the left should be used only as a rough guide to the actual source
brightness.  The solid line shows the observed (but de-redshifted) spectrum
while the dotted line shows the fit described in the text.  Labeled are the
[OIII], H$\beta$ and H$\gamma$ lines as well as the strong Fe II emission 
complexes. 
\label{figure:optical}}
\end{figure*}
Although NLS1 galaxies have become  a topical subject of AGN research,
a general picture explaining the peculiar optical and X-ray
properties has not yet emerged. An important question is how the properties
of NLS1 galaxies depend on the orientation of the sources with respect to
the line of sight. The presence of radio emission and the
associated relativistic jet in radio-loud NLS1 galaxies might provide us 
with a handle on the orientation issue.
Originally it was suggested that NLS1 are the pole-on 
versions of normal Seyfert 1 galaxies. This hypothesis can explain many of
the NLS1 properties, namely the narrower permitted lines, the strong Fe II
emission, the low frequency of warm absorbers and the rapid variability. 
On the other hand the orientation model is in contrast with slab thermal 
Comptonization models
(e.g. Haardt \& Maraschi \cite{har-mar}), which predict softer spectra as
the inclination angle increases. However, if the hot phase responsible for
the Comptonization is contained in discrete regions (blobs) 
above the disk, the X-ray spectral index 
depends mainly on the geometry of the blobs rather than on 
the inclination of the disk (Haardt \cite{har}). 

In this paper we present the X-ray observations both from
BeppoSAX and ROSAT of \object{RX J1702.5+3247}
[${\rm RA}(2000)=17^h02^m31^s,~ {\rm DEC}(2000)=+32^o47\arcmin20\arcsec$; 
z=0.163], together with optical and radio observations.
Throughout the paper we use  a Friedman cosmology with
$H_0 = 70~{\rm kms^{-1}Mpc^{-1}}$ and $q_0 = 0.5$
for the  computation of the K-corrected luminosities.

\section{Radio and optical observations}
\object{RX J1702.5+3247} is a radio-selected source from the FIRST
(Faint Images of the Radio Sky at Twenty centimeters; Becker \cite{beck})
radio survey.  It is an unresolved radio point source with a peak flux of 1.8
mJy.  An optical counterpart was found on the POSS plates with a blue
magnitude of m$_{\rm O}$ = 15.7 and a red magnitude of m$_{\rm E} = $ 15.8.  
This object is a member of
the FIRST Bright Quasar Survey (FBQS; White \cite{fbqs}) and was observed
spectroscopically on June 4, 1997, with GoldCam on the Kitt Peak 2.1-m
telescope.  The dispersion was 1.5 \AA/pix.  All sources with even moderately
broad optical emission lines,
including \object{RX J1702.5+3247}, were classified as quasars in the FBQS.
Our analysis of the optical spectrum (Fig. 1) shows that it can more precisely
be classified as a Narrow Line Seyfert 1 galaxy.

We fitted the optical continuum from 3700 to 6000 \AA ~
using the SPECFIT routine in the IRAF (v.\
2.11.3) data analysis package.  The spectrum was not taken under
photometric conditions, so we restrict our analysis to parameters insensitive
to the absolute flux scale.  The
H$\beta$ line was fit with a Lorentzian profile and has a width of
$\sim$1,400 km s$^{-1}$.  The [OIII] lines were also fit with Gaussian
profiles with widths that were $\sim 650$ km s$^{-1}$.  This is very close to
the resolution of our dataset, so a higher resolution spectrum is required to
determine the true [OIII] line widths.
The Fe II emission was fit using the template from Boroson \& Green
(\cite{BandG}) which was derived from IZw1.  Our spectrum
does not extend far enough to the red to measure the [NII] or H$\alpha$ lines.
In the region localized to the Fe II emission, H$\beta$ and [OIII] lines, we
find the continuum is well parameterized as a power law with a slope of
$\alpha=1.9$
(S$_{\lambda} \propto \lambda^{-\alpha}$).  Similar parameters are reported by
Bade et al.\ (\cite{bade}).  These characteristics taken as a whole exhibit
all the classical features of a NLS1 galaxy: the Fe II emission is strong
compared to a normal Seyfert galaxy and the [OIII] and H$\beta$ emission are
both narrow and strong, although the [O~III] to H$\beta$ flux ratio (0.04) is
remarkably weak in this object.

The redshift of the source is z=0.163\footnote{This redshift differs from
that reported in NED by 0.001 (z$_{\rm NED}=0.164$).  Our redshift is taken from
White \cite{fbqs}, who determined their redshifts by cross-correlating spectra
with a quasar template, effectively taking a global average over all the lines
and features in the observed spectrum.}, which yields a monochromatic radio
luminosity of $1 \times 10^{30}$ erg s$^{-1}$ Hz$^{-1}$.  We also find that the
K-corrected radio-to-optical flux ratio, R$^*$ (following the definition in
Stocke \cite{stocke}), is $1.2$\footnote{Both the monochromatic radio power
and R$^*$ are calculated at 1.4 GHz instead of the standard 5 GHz.  If the
radio spectral index is steeper than zero, both the radio power and R$^*$
will be less than the values reported above}.  Clearly this source is
radio-quiet (as are most presently known NLS1s).  Nevertheless, it is
radio-selected and may possess small-scale, possibly relativistic, jets as has
recently been reported in some Seyfert galaxies (see below).

\section{X-ray observations and data reduction}
\object{RX J1702.5+3247} was first observed with the ROSAT
 (Tr\"umper \cite{tru})
Position Sensitive Proportional Counter (PSPC, Pfeffermann \cite{pfef}) 
during the ROSAT All-Sky Survey (Voges \cite{vog}) on two occasions,
namely on August 18, 1990 and February 17, 1991, with exposures 
of 810 s and 306 s, respectively. 
Further, on August 3, 1993  \object{RX J1702.5+3247} was
 serendipitously observed during a PSPC 
pointed observation of the cluster of galaxies Abell 2241, with an exposure of
$\sim 11$ ks. The analysis of the ROSAT data were performed 
using the EXSAS data analysis package (Zimmermann \cite{zim}). 
For the  pointed observation, the light curve and spectrum
were extracted from a circle with a $300\arcsec$ radius around the source
center; the background was taken from a ring centered on the source with inner
and outer radii of $400\arcsec$ and $500\arcsec$, respectively. 

\object{RX J1702.5+3247} was observed with the BeppoSAX (Boella \cite{boel1})
Narrow Field Instruments: LECS (0.1-10 keV; Parmar \cite{parm}) and MECS
(1.3-10 keV; Boella \cite{boel2}) on March 9-10, 2000 with 
effective exposures of 16.0 ks and 38.3 ks, respectively. The shorter
LECS exposure resulted from the instrument being switched off over the
illuminated Earth. The observations were performed with 2 active MECS units 
(after the failure of MECS1 on May 6, 1997). 
Standard data reduction techniques were employed following the prescription
given by Fiore et al. (1999). LECS and MECS spectra and light curves were
extracted from regions with radii of 8 arcmin and 4 arcmin respectively, in
order to maximize the accumulated  counts at both low and high energies.
Background spectra were extracted from high Galactic latitude ``blank" fields,
whereas for the light curves a source-free region in the field of view was
used. The background subtracted count rates are $(3.83\pm0.22)\times10^{-2}
{~\rm cts~s^{-1}}$ and $(2.15\pm0.09)\times10^{-2}{~\rm cts~s^{-1}}$ for LECS
and MECS respectively.

\section{Temporal analysis}
\begin{figure}
\psfig{figure=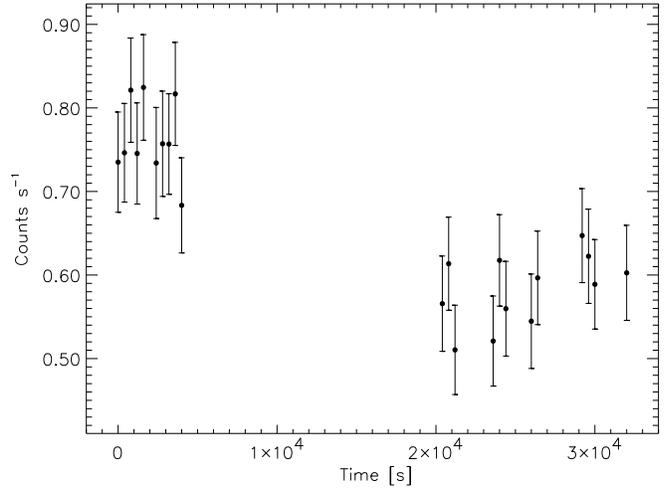,height=6.5cm,width=8.7cm,%
bbllx=64pt,bblly=80pt,bburx=530pt,bbury=715pt,angle=90,clip=}
\caption{ROSAT PSPC light curve on August 3, 1993. Every data point corresponds 
to 400 s integration time.  
\label{figure:PSPClc2}}
\end{figure}
The short exposures during the RASS do not permit a detailed 
temporal analysis. On the other hand, the availability of
two observations taken six months apart allows us to look for 
long term variability.
Indeed, the average count rate changed by a factor $\sim 3$, increasing 
from $0.35\pm0.20$ cts/s on August 18, 1990 to
$1.13\pm0.41$ cts/s on February 17, 1991. There is also evidence for
long term spectral variability: the source spectrum becomes softer
 as the flux increases (Grupe \cite{grupe2}).
Fig.~\ref{figure:PSPClc2} shows the ROSAT PSPC light curve on August 3,
1993 during the pointed observation. An integration time of 400 s per bin 
was used in order to obtain good statistics and avoid count rate
variations due to the ROSAT wobble.  
Fig.~\ref{figure:PSPClc2} shows that \object{RX J1702.5+3247} was moderately
variable during the pointed observation with count rate changes as large as
40\% in a few hours. 

\begin{figure}
\psfig{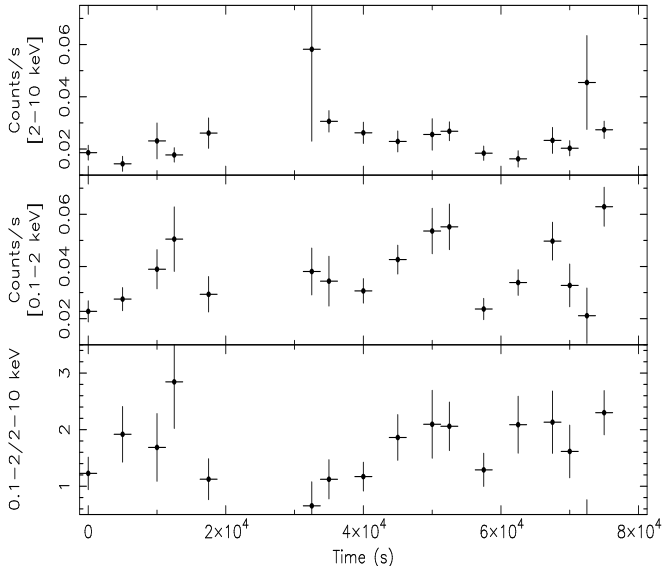}
\caption{MECS (top panel), LECS (middle panel) and soft-to-hard counts ratio 
(bottom panel)
of \object{RX J1702.5+3247} on March 9-10, 2000, with time binning of 2500 s.
\label{figure:le-me-ha}}
\end{figure}

In Fig.~\ref{figure:le-me-ha} the LECS and MECS light curves with time
binning of 2500 s are shown, together with the corresponding
[0.1-2 keV]/[2-10 keV] soft-to-hard counts ratios.
 While the MECS light curve
is basically consistent with the hypothesis of constant flux (the $\chi^2$
probability that the flux is constant is p$\,\sim0.9$), variability up to a factor  of 
2 on time scales of hours is detected in the LECS light curve with a
high level of confidence (${\rm p}<3\times10^{-4}$). There is also evidence of 
spectral variability. Since the flux is dominated by the soft photons,
the softness ratio follows the low energy light curve, and the 
spectrum softens when the total flux increases. 

\begin{figure}
\psfig{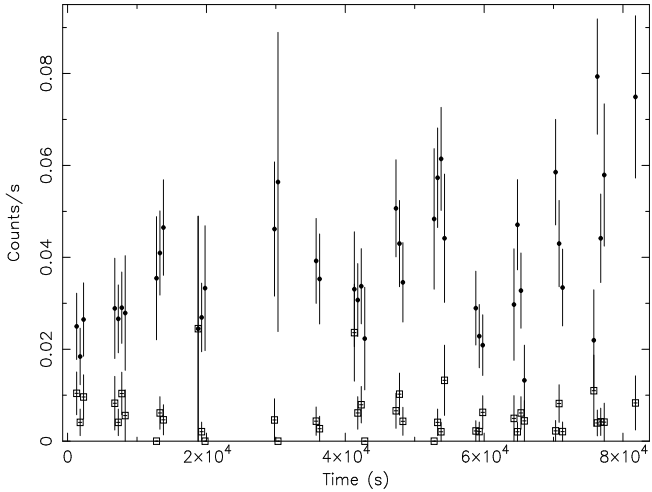}
\caption{LECS light curve. Every data point corresponds to 500 s integration
time. Filled circles and open squares represent the 
source and the background count rates, respectively.
\label{figure:le-bkg-500}}
\end{figure}

\begin{figure}
\psfig{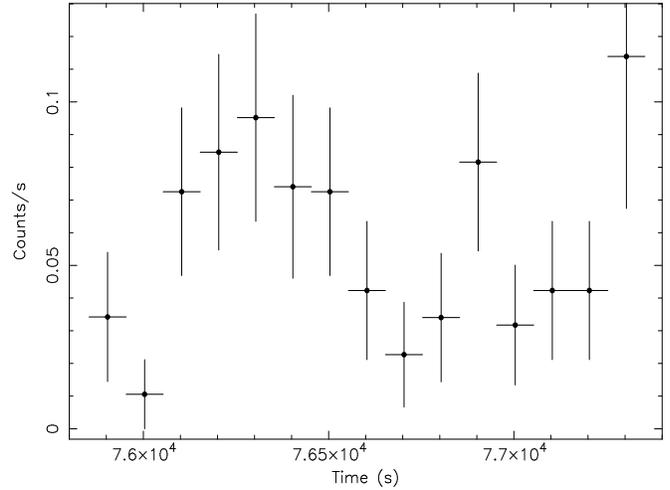}
\caption{LECS light curve of orbit 13. Every data point corresponds to 100 s 
integration time.
\label{figure:lecs100-13}}
\end{figure}

A characteristic
property exhibited by several NLS1 galaxies is the presence of large flares on
very short time scales. 
Fig.~\ref{figure:le-bkg-500}, which shows the LECS light curve with 500 s
binning, reveals the presence of significant variability
(p$\,<0.15$) on shorter time scales, namely within the individual orbits, 
while the background  count rate remains basically constant. In particular,
around t=$7.5\times 10^4$ s, the source count rate varies by a factor of 4 in 
500 s. We therefore investigated the LECS light curve with even 
smaller binning (100 s), and found that in at least four orbits there
is evidence of strong variability on very short time scales. For brevity
we concentrate on orbit 13 (using the data between $t=7.59\times 10^4$ s  
and $t=7.67\times 10^4$ s after the beginning of the observation), 
which contains the largest number of data
points during a flare (see Fig.~\ref{figure:lecs100-13}).
We performed a linear least square fit to determine the count rate variations
during the rise and the decline of the flare 
 and found ${\rm \Delta rate/\Delta t}=(2.28\pm0.75)
\times10^{-4}{\rm ~cts~s^{-2}}$ and ${\rm \Delta rate/\Delta t}=(2.06\pm0.27)
\times10^{-4}{\rm ~cts~s^{-2}}$, respectively. 
We calculated the corresponding luminosity
variations  in the rest frame of \object{RX J1702.5+3247} 
by assuming a power law spectral model with the best fit parameters
(discussed in Sec. 5) of the 0.1-2 keV LECS
spectrum. We obtained
${\rm \Delta L_{\rm 0.1-2 keV}/\Delta t}=(5.1\pm1.7)\times10^{42}
{\rm erg~s^{-2}}$ for the rise and ${\rm \Delta L_{\rm 0.1-2 keV}/\Delta t}=
(4.6\pm0.6)\times10^{42}{\rm erg~s^{-2}}$ for the decay, which are 
extremely high values, even for NLS1 galaxies. 
We are aware that the small number of data points
yield a less than robust statistical significance. 
On the other hand, the presence of 
similar short flares in different orbits supports the hypothesis of strong 
variability.
For a firm confirmation we have to await further observations
from the new X-ray missions such as XMM-Newton, with sensitive, 
high time resolution 
instruments and the possibility of long continuous observations.
 
The luminosity variations can be used to estimate the lower limit of the 
radiative efficiency: $\eta > 4.8\times 10^{-43}{\rm\Delta}L/{\rm\Delta}t$ 
(Fabian \cite{fabi}).
Straightforward application of this relation  gives $\eta > 2$  
for both the rise and the decay.  This suggests 
that some assumptions used in the calculation of the efficiency limit
must be relaxed, allowing uniform radiation release
and  relativistic effects in the vicinity of the black hole
(e.g. Brandt \cite{brand2}). 
Interestingly, a similar extreme value for the efficiency 
was derived by Remillard et al. (1991) from the Ginga 2--10 keV observation  
of \object{PKS~0558--504}, a radio-loud NLS1 galaxy. 

We compared the ROSAT PSPC count rate with the LECS, by using W3PIMMS
with a power law model and the best fit parameters ($\Gamma=3.37$,
$N_{\rm H}=2.47\times 10^{20}{~\rm cm^{-2}}$) of the PSPC spectrum during
the pointed observation. The average count rate of $6.6 \times 10^{-1} 
{~\rm cts~ s^{-1}}$, found with 
the PSPC on August 3, 1993, corresponds to  $3.85\times 10^{-2}
{~\rm cts~ s^{-1}}$ 
in the 0.1-2 keV energy band of the LECS, which is consistent with
the average rate of $3.83\times 10^{-2}{~\rm cts~ s^{-1}} $  measured
in March 2000.

\section{Spectral analysis}

\subsection{Spectral fitting}
\begin{figure}
\psfig{figure=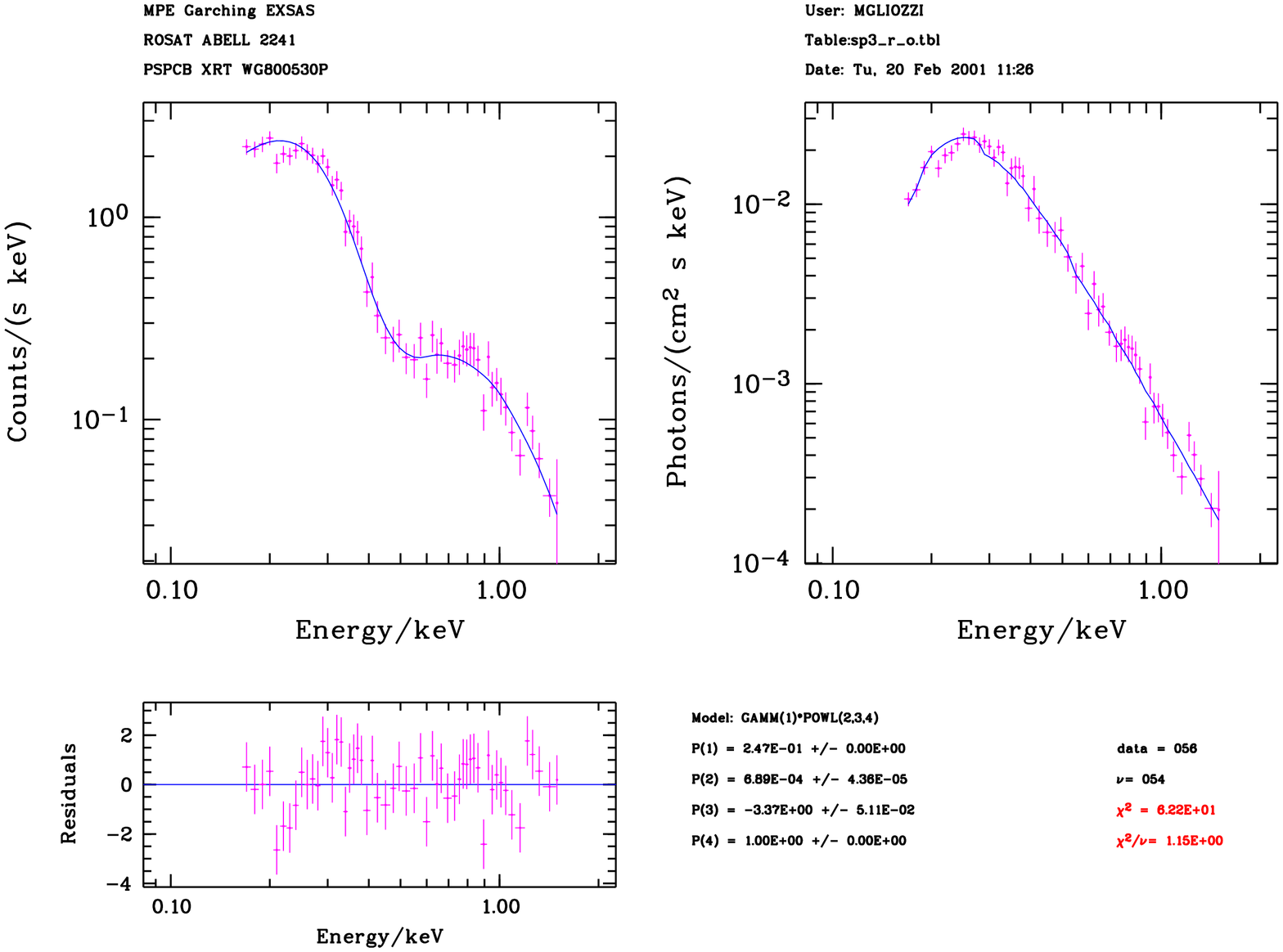,height=4.4truecm,width=8.5truecm,angle=-90,%
bbllx=340pt,bblly=50pt,bburx=75pt,bbury=400pt,clip=}
\psfig{figure=pow3a.ps,height=2.6truecm,width=8.5truecm,angle=-90,%
bbllx=551pt,bblly=50pt,bburx=414pt,bbury=400pt,clip=}
\caption[]{Single power law fit of the ROSAT PSPC pointed observation  
with $N_{\rm H}$ fixed to the Galactic value for \object{RX J1702.5+3247}; 
top panel the folded spectrum, bottom panel the residuals.
\label{figure:pow3}}
\end{figure}

Spectra from the ROSAT data were produced using the standard procedures from the
EXSAS analysis package. Source counts were binned to give a minimum 
signal to noise ratio of $\geq$ 5.
The poor photon statistics of the spectrum of \object{RX J1702.5+3247}, 
obtained from the RASS data, do not allow sophisticated fits. 
A power law plus absorption  model
 reveals that the intrinsic absorption in the source is 
negligible and therefore we fixed the absorption at the Galactic value
($N_{\rm H}=2.47\times 10^{20}{~\rm cm^{-2}}$; Dickey \& Lockman \cite{di-lo}).
The resulting best fit parameters are $\Gamma=3.13\pm0.14$ and $\chi^2=9.4$ 
(12 d.o.f.). Fig.~\ref{figure:pow3} shows that also during the PSPC 
pointed observations a single power law fit 
with column density fixed to the Galactic value is an acceptable description
($\chi^2=62.2$ for 54 d.o.f.) for the soft spectrum.
The results of the spectral fits assuming Galactic absorption 
are summarized in Table 1.
On the basis of the best fit model, the unabsorbed 0.1-2.4 keV flux is
$1.8\times 10^{-11} {\rm ~erg~cm^{-2}~s^{-1}}$, corresponding to a soft
X-ray luminosity of $1.4\times 10^{45} {\rm~erg~s^{-1}}$ (on August 3, 1993).

For the spectral analysis of the BeppoSAX data we used the XSPEC 10 software 
package and the latest
release of the response matrices. LECS and MECS spectra were rebinned in order
to have at least 20 counts per energy channel. 
Due to the slight mis-match in the
cross-calibration of the different detectors, it has become necessary to
include multiplicative factors for the normalization of the fitted models. The
correct absolute flux normalization is provided by the MECS and the expected
value of these constant factors is well known (between 0.7 and 1) and
does not constitute an additional source of uncertainty (Fiore \cite{fior2}).
In the following, all quoted errors correspond to 90\% confidence levels.

\begin{table} 
\caption{Single power law fits with fixed Galactic absorption.}
\begin{center}
\begin{tabular}{lcll}
\hline
\hline
\noalign{\smallskip}
Instrument & $\Gamma$ & $\chi^2/{\rm d.o.f.}$ & $\chi^2_{\rm red}$\\
\noalign{\smallskip}       
\hline
\hline
\noalign{\smallskip}
\noalign{\smallskip}
PSPC (RASS)& $3.13\pm0.14$ & 9.4/12 & 0.78\\
\noalign{\smallskip}
PSPC (pointed)&$3.37\pm0.05$ & 62.2/54 & 1.15 \\
\noalign{\smallskip}
\hline
\noalign{\smallskip}
LECS & $2.94\pm0.10$ & 40.02/37 & 1.14\\
\noalign{\smallskip}
\hline
\noalign{\smallskip}
MECS & $2.43\pm0.20$ & 48.08/35 & 1.37\\
\noalign{\smallskip} 
\hline
\end{tabular}
\end{center}
\end{table}
 \begin{figure}
\psfig{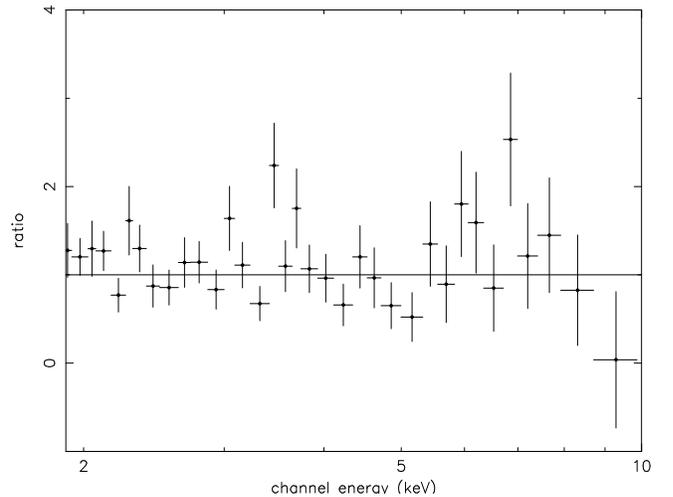}
\caption{Data to model ratio for the power law fit to the 2-10 keV
MECS spectrum.
\label{figure:po2-10a.ps}}
\end{figure}
A summary of the individual
LECS and MECS datasets fitted with a single power law plus Galactic absorption
is reported in Table 1. Leaving the column density free to vary, one gets 
similar results with a best fit value of $N_{\rm H}=(1.85\pm0.8)\times
10^{20}~{\rm cm^{-2}}$, consistent with the Galactic value. The relatively
poor fit of the MECS spectrum can be ascribed to the presence of line-like
excess emission around 7 keV. 

\begin{figure}
\psfig{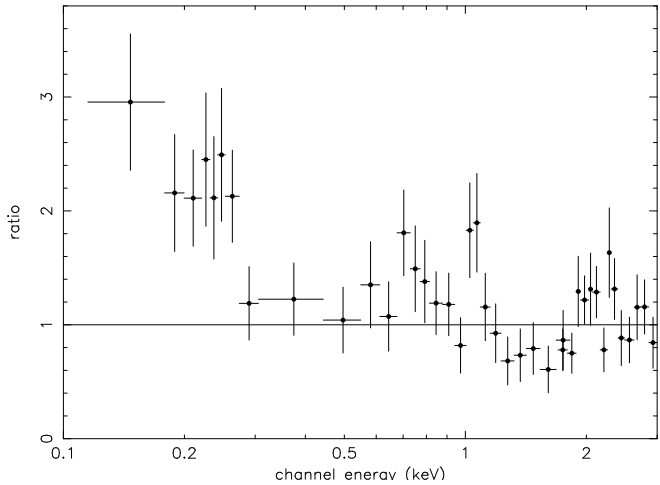}
\caption{Data to model ratio given a power law fit to the 2-10 keV
spectrum.
\label{figure:le-powme}}
\end{figure}

By extrapolating the power law model fitted in the 2-10 keV band
down to 0.1 keV (Fig.~\ref{figure:le-powme}), a soft component is 
revealed, but only at very low energies below 0.3 keV.

A simultaneous fit  of a power law model with fixed  Galactic absorption 
to the LECS (restricted to the range 0.1--2 keV, to avoid the large errors at
higher energies) and MECS data resulted in  a slope of  
$\Gamma=2.81\pm0.11$ and a relatively poor $\chi^2=87.88$ (61 d.o.f.).
A statistically significant improvement of the fit
($\Delta\chi^2=5.68$, significant at more than 90\% confidence
for two additional free parameters) is achieved by
adding a Gaussian line profile to the power law model. In order to determine 
the line parameters, given the quality of the data, the energy centroid
of the Gaussian was constrained to lie in the range 5.5--7.5 keV 
and its width was fixed at 0.1 keV. The resulting line is 
centered at  $7.11_{-0.28}^{+0.32}$ keV in the rest frame, and has an equivalent width of
$804_{-511}^{+856}$ eV. Such a high value for the equivalent width
is not uncommon among NLS1 galaxies: similar values were reported for 
\object{Ton~S~180} (${\rm EW}=507\pm247$ eV, Comastri \cite{coma1})
and \object{PG~1244+026} (${\rm EW}=594$ eV, Ballantyne \cite{balla}).
We also tried to fit the data with a power law plus a relativistic
disk line (Laor 1991), however, neither a significant improvement of the fit 
nor a more accurate determination of the line parameters was achieved.

\begin{figure}
\psfig{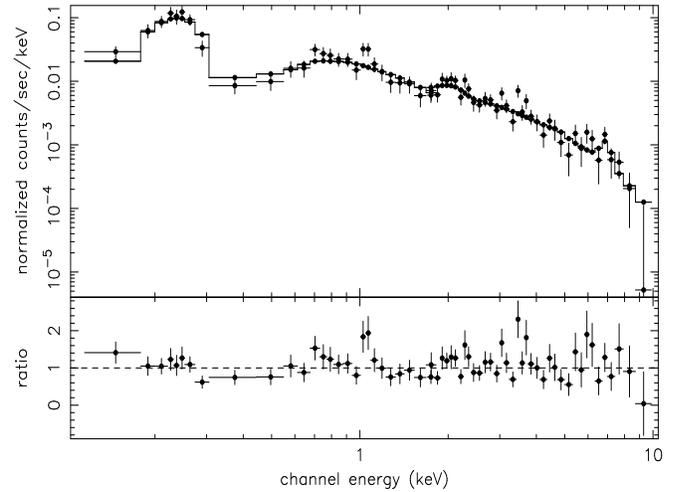}
\caption{LECS and MECS spectrum of \object{RX J1702.5+3247} with a 
broken power law model plus a Gaussian line superimposed. 
\label{figure:bkn-ga1}}
\end{figure}

We therefore tried more complex two-component models which
are currently used to fit the broad band X-ray spectra of NLS1 galaxies (e.g.
Leighly \cite{leig2}): a blackbody model superimposed  on a simple
power law, a bremsstrahlung model plus a power law, and a broken power law. Only
the latter gave a statistically significant improvement of the fit: the
difference in $\chi^2$ is significant at more than 99.9\% confidence, according
to an F--test. 
An even better fit (at confidence level between 68\% and 90\%) is 
obtained by adding a Gaussian line. The resulting fit, superimposed on the 
data, is shown in Fig.~\ref{figure:bkn-ga1}.  All LECS+MECS 
joint fits are summarized in Table 2.
On the basis of the best fit model, the unabsorbed 0.1-2.4 keV flux is
$1.18\times 10^{-11} {\rm ~erg~cm^{-2}~s^{-1}}$, corresponding to a soft
X-ray luminosity of $8.5\times 10^{44} {\rm~erg~s^{-1}}$.
The 2.4-10 keV flux is $0.9
\times 10^{-12} {\rm ~erg~cm^{-2}~s^{-1}}$, corresponding to a hard X-ray
luminosity of $6.4\times 10^{43} {\rm~erg~s^{-1}}$ 
is roughly a factor 13 lower than the soft one.

\begin{table*} 
\caption{LECS+MECS joint fits to the 0.1-10 keV continuum.  All values
quoted are in the rest frame.}
\begin{center}
\begin{tabular}{llllllll}
\hline
\noalign{\smallskip}
Model & $\Gamma_1$ & $E_{\rm br} (keV)$ & $\Gamma_2$ & $E_{\rm line} (keV)$ &
 EW (eV)& $\chi^2/{\rm d.o.f.}$ & $\chi^2_{\rm red}$\\
\noalign{\smallskip}       
\hline
\hline
\noalign{\smallskip}
\noalign{\smallskip}
pow & $2.81^{+0.11}_{-0.11}$ &  &  &   & &87.88/61 & 1.44  \\
\noalign{\smallskip}
\hline
\noalign{\smallskip}
po+ga & $2.83^{+0.11}_{-0.11}$ &  & &$7.11^{+0.32}_{-0.28}$ & $804^{+856}_{-511}$&
82.20/59 & 1.39  \\
\noalign{\smallskip}
\hline
\noalign{\smallskip}
bkn & $2.93^{+0.14}_{-0.13}$ & $1.7^{+0.8}_{-0.6}$ & $2.45^{+0.19}_{-0.22}$ & & &
77.09/59 & 1.30  \\
\noalign{\smallskip}
\hline
\noalign{\smallskip}
bkn+ga & $2.92^{+0.14}_{-0.12}$ & $1.62^{+0.82}_{-0.57}$ & $2.50^{+0.20}_{-0.21}$  & $8.08^{+0.55}_{-1.18}$ 
 & $808^{+682}_{-758}$  &73.94/57 & 1.30  \\
\noalign{\smallskip} 
\hline
\noalign{\smallskip}
\noalign{\smallskip}
\noalign{\smallskip}
\hline
\noalign{\smallskip}
Model &  $\Gamma$ & $\log \xi$ & R &  &
&$\chi^2/{\rm d.o.f.}$ & $\chi^2_{\rm red}$\\
\noalign{\smallskip}       
\hline
\hline
\noalign{\smallskip}
refl disk & $2.56^{+0.13}_{-0.12}$ & $3.40^{+0.35}_{-0.31}$ &
$0.94^{+0.86}_{-0.57}$ & & & 76.54/60 & 1.27\\
\noalign{\smallskip}
\hline
\end{tabular}
\end{center}
\end{table*}

\subsection{Ionized disk models}
Stimulated by very recent spectral results on ASCA data of NLS1 galaxies 
(Ballantyne \cite{balla}), where the ionized reflection model 
(Ross \& Fabian \cite{ros-fab}, Ross et al. \cite{ros2}) 
proved to provide excellent fits to the data, 
we performed a similar analysis on the SAX data. In this model, the
accretion disk is approximated as a slab
of gas with constant density ($n_{\rm H}=10^{15}{\rm~cm^{-3}}$) and solar
abundances, which is illuminated by an X-ray flux $F_{\rm x}$, described by
a power law with photon index $\Gamma$. The reflected spectrum is multiplied
by $R$, the reflected fraction, and then added to the incident spectrum to give
the resulting observed spectrum. The structure of the reflected spectrum is
determined by the ionization parameter $\xi$, which varies as a
function of the incident flux. Recently, Nayakshin \cite{naya} presented
more sophisticated ionized reflection models, where the density is determined 
from hydrostatic balance, which is solved simultaneously with ionization balance
and radiative transfer. The authors point out that, due to a thermal
instability, only a very thin layer at the top of the slab is highly ionized,
and the reflected spectrum is dominated by the neutral material underneath.
However, in case of a steep ($\Gamma>2$) incident spectrum, as typical
for NLS1 galaxies, the Nayakshin et al. model predicts the presence of 
an ionized region of substantial optical depth and therefore ionized features
superimposed on the emerging X-ray spectrum. 
Thus, despite the simplifying assumption of constant density, 
the use of the Ross \& Fabian ionized reflection model is justified
in the case of the steep spectrum NLS1 galaxy \object{RX J1702.5+3247}.
Further it is
preferable to the alternative ionized disk model (PEXRIV) adapted in the XSPEC 
spectral fitting procedures, for its capacity to fit simultaneously the
continuum and the superimposed lines. 
\begin{figure}
\psfig{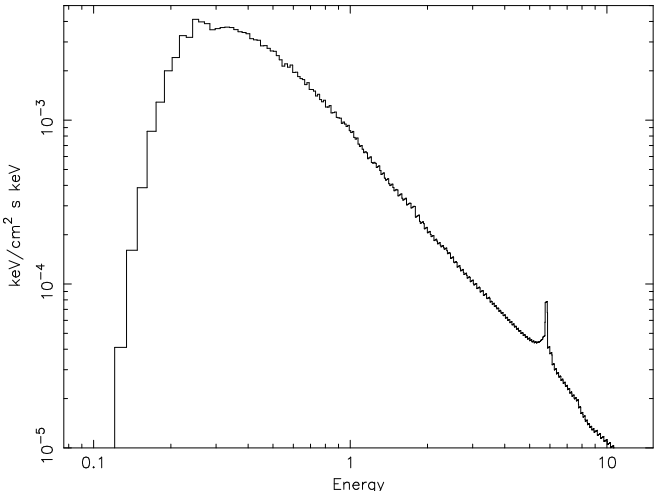}
\caption{Best fit ionized disk model for \object{RX J1702.5+3247}. The energy
is in the observer frame.
\label{figure:io-model}}
\end{figure}
Fig.~\ref{figure:io-model} shows the best fit ionized disk model ($\Gamma=2.56$,
$\log\xi=3.4$, $R=0.94$). The fit parameters are consistent with those
found for \object{Ton S 180} by Ballantyne \cite{balla}, although the reflection
fraction $R$ of \object{RX J1702.5+3247} is poorly constrained.)
The most prominent feature is the Fe K$\alpha$ line
at 5.9 keV in the observer rest frame, which corresponds to a rest frame energy
of 6.9 keV, in full agreement with the fluorescent emission from H-like Fe
ions. The ionized disk model fits the Fe line and the
continuum simultaneously and thus determines the Fe K$\alpha$ centroid
energy, but no further information on the line parameters is provided. 
The ratio between data and model, shown in Fig.~\ref{figure:io-ratio1},
demonstrates that the ionized disk model gives an acceptable representation of 
the spectrum over the entire energy range. 
A comparison between the ionized disk model 
and the broken power law plus a Gaussian line indicates that the latter
is not preferable, since a $\Delta\chi^2=2.6$ is significant at less
than 68\% confidence level for three additional free parameters.
Moreover, for both models we calculated the probability to find a better
reduced $\chi^2$ given the respective $\chi^2$ and degrees of freedom. The 
result shows that the ionized disk model and the broken power law plus a 
Gaussian line cannot be discriminate 
at a confidence level of 99\%. 

\begin{figure}
\psfig{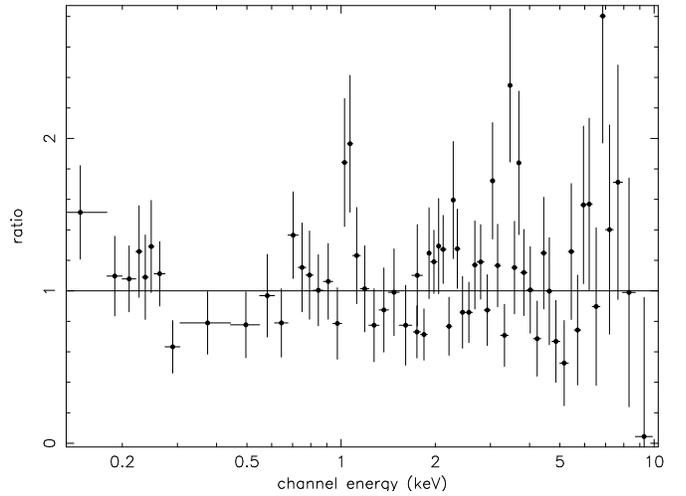}
\caption{Data to model ratio for an ionized disk model.
The energy is in the observer frame.
\label{figure:io-ratio1}}
\end{figure}

\section{Discussion}
The X-ray properties of \object{RX J1702.5+3247} are remarkable, both in
the temporal and spectral behaviour. The timing analysis
revealed the presence of significant flux variations on very small time scales,
which lead to an extreme value ($\eta>2$) for the efficiency in the conversion 
of gravitational potential energy into X-ray emission. 
Although large amplitude variability
on short time scales is a common feature among NLS1 galaxies, up to now
a similar extreme behaviour was observed only in \object{PKS~0558-504}
(Remillard \cite{remi}), and was attributed to relativistic beaming
occurring in the jet, given the radio-loud nature of this source.
\object{RX J1702.5+3247} was detected in the FIRST VLA radio survey, but
is formally defined as a radio--quiet source, on the basis of
the radio-to-optical flux ratio R. 
On the other hand, the extreme value of $\eta$ indicates the presence of 
beaming effects, which might be naturally explained by
the existence of a relativistic jet albeit here in a radio--quiet object.
This scenario has already been
proposed for radio-intermediate quasars on the basis of VLBI observations 
and brightness temperature calculations (e.g. Falke \cite{falke}).
Such a hypothesis finds further support in recent VLBA radio observations of 
Seyfert galaxies, where sub-relativistic jets on parsec-scale
were detected (e.g. Ulvestad \cite{ulve2}) and in one case 
(\object{III Zw 2}; Brunthaler \cite{brunt}) even superluminal motion was 
measured. 
An alternative explanation is that the fast variability is
related to strong relativistic effects occurring in the accretion disk
seen under large viewing angles (namely an edge--on disk; see e.g. 
Cunningham \cite{cunni}). However this interpretation seems to be in conflict
with the large equivalent width of
the iron K$\alpha$ fluorescent line, which is expected to decrease
rapidly with increasing inclination angles due to relativistic effects
(Matt \cite{matt}).

The analysis of the spectral properties of \object{RX J1702.5+3247} suggests
(at a confidence level between 68\% and 90\%) the presence of a 
{$\rm K_{\alpha}$} fluorescent iron emission line. The centroid energy of this
line is poorly determined, but is inconsistent with the bulk of the line being 
produced by iron in a ionization state lower than He-like.
It should be noted that high energy iron K$\alpha$ features have already been
detected in two other NLS1 galaxies, Ton S 180 and Ark 564, observed  with 
BeppoSAX (Comastri et al. 1998, 2001), by fitting ionized reflection and 
absorption models, whose limits have been outlined by Ballantyne et al. (2001). 
The last authors successfully fit the ASCA data of Ton S 180 and Ark 564, 
among other NLS1 spectra, with the Ross \& Fabian reflection model, but without
finding any emission line consistent with highly ionized iron.
Therefore \object{RX J1702.5+3247} is the first object where evidence for
a strongly ionized accretion disk is twofold. First, a Gaussian line at 
energy consistent with that of iron in highly ionized state
is required in any spectral model
for a significant improvement of the fit. Furthermore, the joint LECS+MECS 
spectrum
is well fitted by the Ross \& Fabian ionized disk reflection model
with a high ionization parameter ($\xi \sim 2500$).

The present results fit fairly well with the hypothesis that NLS1 have higher 
accretion rates, relative to the Eddington value, compared to normal broad
line Seyfert galaxies. In fact, high accretion rates are accompanied by
strong ionization of the disk surface layers. According to the 
hypothesis of higher accretion rates and consequently smaller black hole
masses (provided that the energy conversion efficiency is the same), the
steep soft excess is explained  by the shift of the accretion disk spectrum
into the soft X-ray band. The strong flux of soft photons could lead to a
strong Compton cooling in the corona and thus to a steep hard tail.
Despite the rather steep soft photon index
($\Gamma=3.37$ in the ROSAT observation), \object{RX J1702.5+3247}
shows only a weak soft excess, confined to the very soft part ($<0.3$ keV) of
the X-ray spectrum. However, the lack or weakness of the soft component 
is not uncommon among NLS1 galaxies, as noted by Leighly (\cite{leig2}), who
also found that the soft excess strength is related to the differences in the
relative normalizations of the soft excess component and the power law, rather
than to the difference in the slopes of the soft and hard band power laws.

\begin{acknowledgements} 
We thank the anonymous referee for the useful comments and suggestions that
improved the paper.    
MG  acknowledges support from
the European Commission under contract number ERBFMRX-CT98-0195 
(TMR network ``Accretion onto black holes, compact stars and protostars").
SALM acknowledges support from NSF grant AST-98-02791 and from the Institute
of Geophysics and Planetary Physics (operated under the auspices of the U.S.
Department of Energy by Lawrence Livermore National Laboratory under contract
No.\ W-7405-Eng-48)

\end{acknowledgements}

\end{document}